\documentstyle[11pt]{article}

\newcommand{\be}{\begin{equation}}
\newcommand{\ee}{\end{equation}}
\newcommand{\bn}{\begin{eqnarray}}
\newcommand{\en}{\end{eqnarray}}
\newcommand{\bd}{\begin{displaymath}}
\newcommand{\ed}{\end{displaymath}}

\begin{document}
\title{COSMIC STRING CREATED \\ FROM VACUUM FLUCTUATION}
\author{Arkadii Popov \thanks{E-mail: popov@kspu.ksu.ras.ru}
\\Department of Geometry, Kazan State Pedagogical University,\\
Mezhlauk str., 1, Kazan 420021, RUSSIA}

\date{}
\maketitle

\begin{abstract}
The possibility of the cosmic string creation by the vacuum fluctuations of
quantum fields in the self-consistent semiclassical theory of gravity
is discussed. We use the approximate method for obtaining vacuum
expectation value of the renormalized stress-energy tensor of conformally
invariant quantum fields in static cylindrically symmetric spacetimes.
We have obtained the particular solutions of Einstein equations
for the different boundary conditions at the cylinder symmetry axis.
\end{abstract}

There is very interesting question: can the vacuum fluctuation of quantum
fields to be the single source of curvature or nontrivial topological
structure of spacetime? One may find the solution of this problem in the
theory of quantum gravity. But a fully satisfactory theory of quantum
gravity does not yet exist. As a first step on the way of solution of this
problem one may to use the self-consistent semiclassical theory of gravity.
This theory sets the classical Einstein tensor equal to the expectation
value of the stress-energy tensor operator of the quantized matter fields
present (throughout we use units such that $c= \hbar =G=1$)
\be
\label{1} G_{\mu \nu} = 8\pi \langle T_{\mu \nu} \rangle.
\ee
The primary difficulty in the theory of semiclassical gravity is that
the effects of the quantized gravitational field are ignored.
A secondary difficulty of semiclassical gravity is that
$ \langle T_{\mu \nu} \rangle $ depends strongly on the metric  and
is generally difficult to calculate. Most calculations of
$ \langle T_{\mu \nu} \rangle $ have been made on fixed classical background.
These results can not be used in self-consistent semiclassical theory.
There have been approximate calculations made for conformally coupled
fields in any static spherically symmetric Einstein spacetime \cite
{Page,Brown}.
The analytical approximations to $ \langle T_{\mu \nu} \rangle $
have been obtained by Anderson, Hiscock and Samuel \cite{Anderson} for
massless or massive scalar fields with an arbitrary coupling $\xi$ to
scalar curvature in static spherically symmetric spacetimes.
In this work we use the Killing ansatz giving approximate
vacuum expectation value of the stress-energy tensor $ \langle T_{\mu \nu}
\rangle $ for conformally coupled massless field
in static spacetimes which was obtained by Frolov and Zelnikov
\cite{Frolov}.
The Killing ansatz is constructing as an
approximate expression for the vacuum expectation value
$\langle T_{\mu \nu }\rangle $ using the Riemann tensor, the Killing
vector and their covariant derivatives. The tensor $\langle T_{\mu \nu }
\rangle $ is constructed so that the conserva\-tion law $\langle
T_{\mu}^ {\nu } \rangle _{;\nu} =0 $ and the conformal trace anomaly  are
fulfilled. In addition a correct scaling transformation behaviour of
$\langle T_{\mu \nu }\rangle $ is necessary. Then the following
expression can be obtained for $\langle T_{\mu \nu }\rangle $ :
\begin{equation}
\label{2} \langle T_{\mu \nu }\rangle =
\langle T_{\mu \nu } (g_{\mu \nu },\xi ^\mu ,\ q_1,\ q_2) \rangle ,
\end{equation}
where ${g_{\mu \nu }}$ is the metric tensor, $\xi ^\mu $ is the Killing
vector of static spacetime, $q_1$ and $q_2$ are the arbitrary constants
of ansatz.

The metric for the general static  cylindrically symmetric spacetime is
\be
\label{metric} ds^2=-e^{2F}dt^2+b^{2}dz^2+dr^2+c^{2}d \varphi ^2,
\ee
where $F, \ b$ and $c$ are functions of the proper distance $r$.

We restrict our consideration to the special case
\be
F=const.
\ee
>From this assumption it is easy to obtained the following expression
for $\langle T_{\mu \nu }\rangle $ \cite{Sushkov}:

$$
T_{\mu \nu }=2\left[ \alpha (F-F_0)+q_1\right] \left[ \frac 23R_{;\mu \nu
}-2R_{\mu \nu ;\sigma }^{;\sigma }-\frac{14}3RR_{\mu \nu }+8R_{\mu \sigma
}R_\nu ^\sigma
\right.
$$
$$
\left.
+g_{\mu \nu }\left( \frac 13R_{;\sigma }^{;\sigma
}-3R_{\alpha \beta }R^{\alpha \beta }+\frac 53R^2\right)
+\xi _{\mu \nu }\left( 4R_{\alpha \beta }R^{\alpha \beta
}-2R^2\right) \right]
 +\alpha \xi _{\mu \nu }\left( \frac 23R_{;\sigma
}^{;\sigma } \right.
$$
\be
\left.
+2R_{\alpha \beta }R^{\alpha \beta }-\frac 23R^2\right)
+q_2\left[ \frac 13R_{;\mu \nu }-RR\mu \nu +g_{\mu \nu }\left( -\frac
13R_{;\sigma }^{;\sigma }+\frac 1{12}R^2\right)
+ \xi _{\mu \nu }R_{;\sigma }^{;\sigma }\right],
\ee
where the notations are used :
\be
\xi^{\mu}(e^{-F_0},0,0,0), \quad
\xi^2=\xi^{\sigma} \xi _{\sigma} , \quad \xi _{\mu \nu}=\xi_{\mu} \xi_{\nu}/
\xi^2,
\ee
\be
\alpha =\frac 1{2^9\cdot 45\pi ^2}\left[ 12h\left( 0\right) +18h\left( \frac
12\right) +72h\left( 1\right) \right] ,\
\ee
$h\left( s\right) $ is the number of helisity states for the conformal
massless field of a spin $s$.

Four nontrivial Einstein equations for this case are given in Appendix.
This equation set are compatible only in the case
\be
q_2=0, \quad 2 \lbrack \alpha (F-F_0) + q_1 \rbrack = - \alpha
\ee
and independent equations of this set can be written in form
$$
\frac{b^{\prime \prime }}{2b}+\frac{c^{\prime \prime }}c+\frac{b^{\prime
}c^{\prime }}{2bc}=8\pi \alpha \left( -\frac{b^{\prime \prime \prime \prime }%
}b+\frac{5b^{\prime \prime \prime }b^{\prime }}{3b^2}-2\frac{b^{\prime
\prime \prime }c^{\prime }}{bc}-\frac{c^{\prime \prime \prime }b^{\prime }}{%
3bc}+\frac{5b^{\prime \prime 2}}{3b^2}-\frac{4b^{\prime \prime }c^{\prime
\prime }}{3bc}\right.
$$
\be
\label{eq1}
\left. -\frac{5b^{\prime \prime }b^{\prime 2}}{3b^3}+\frac{b^{\prime \prime
}b^{\prime }c^{\prime }}{b^2c}+\frac{b^{\prime \prime }c^{\prime 2}}{bc^2}-
\frac{c^{\prime \prime }b^{\prime 2}}{3b^2c}+2\frac{c^{\prime \prime
}b^{\prime }c^{\prime }}{bc^2}+\frac{b^{\prime 3}c}{3b^3c}-\frac{b^{\prime
}c^{\prime 3}}{bc^3}\right)
\ee
$$
\frac{b^{\prime \prime }}b+\frac{c^{\prime \prime }}{2c}+\frac{b^{\prime
}c^{\prime }}{2bc}=8\pi \alpha \left( -\frac{c^{\prime \prime \prime \prime }%
}c-\frac{b^{\prime \prime \prime }c^{\prime }}{3bc}-2\frac{c^{\prime \prime
\prime }b^{\prime }}{bc}+\frac{5c^{\prime \prime \prime }c^{\prime }}{3c^2}-
\frac{4b^{\prime \prime }c^{\prime \prime }}{3bc}+\frac{5c^{\prime \prime 2}
}{3c^2}\right.
$$
\be
\label{eq2}
\left. +2\frac{b^{\prime \prime }b^{\prime }c^{\prime }}{b^2c}-\frac{%
b^{\prime \prime }c^{\prime 2}}{3bc^2}+\frac{c^{\prime \prime }b^{\prime 2}}{%
b^2c}+\frac{c^{\prime \prime }b^{\prime }c^{\prime }}{bc^2}-\frac{5c^{\prime
\prime }c^{\prime 2}}{3c^3}-\frac{b^{\prime 3}c^{\prime }}{b^3c}+\frac{%
b^{\prime }c^{\prime 3}}{bc^3}\right)
\ee
The solution without source at the symmetry axis must satisfy the boundary
conditions:
\be
b(r)=b_0+\frac{b_0^{\prime \prime}}{2!} r^2+\frac{b_0^{\prime \prime
\prime \prime}}{4!} r^4+0(r^5) \ \ \ \mbox{for}
 \ \ r \rightarrow 0,
\ee
\be
c(r)=c_0^{\prime}r+\frac{c_0^{\prime \prime \prime}}{3!} r^3+0(r^5)
 \ \ \ \mbox{for}
 \ \ r \rightarrow 0,
\ee
where
\be
c_0^{\prime}=1, \ b_0^{\prime \prime \prime \prime}=b_0 \Big[
\frac{b_0^{\prime \prime 2}}{b_0^2}-\frac3{8(8 \pi \alpha)}
\Big(\frac{b_0^{\prime \prime}}{b_0}+\frac{c_0^{\prime \prime \prime}}
{c_0^{\prime}} \Big) \Big],
\ee
$b_0, b_0^{\prime \prime}, c_0^{\prime \prime \prime}$ are arbitrary
constants. These expressions can be obtained from equations (\ref{a1})
-(\ref{a4})
by substitution the functions $b(r)$ and $c(r)$ power series of $r$.

The equations (\ref{eq1}) and (\ref{eq2}) were solved by numerical methods
for the scalar field
$\left( \displaystyle \alpha=\frac{12}{2^9\cdot 45\pi ^2} \right)$.
The results are demonstrated in figures:

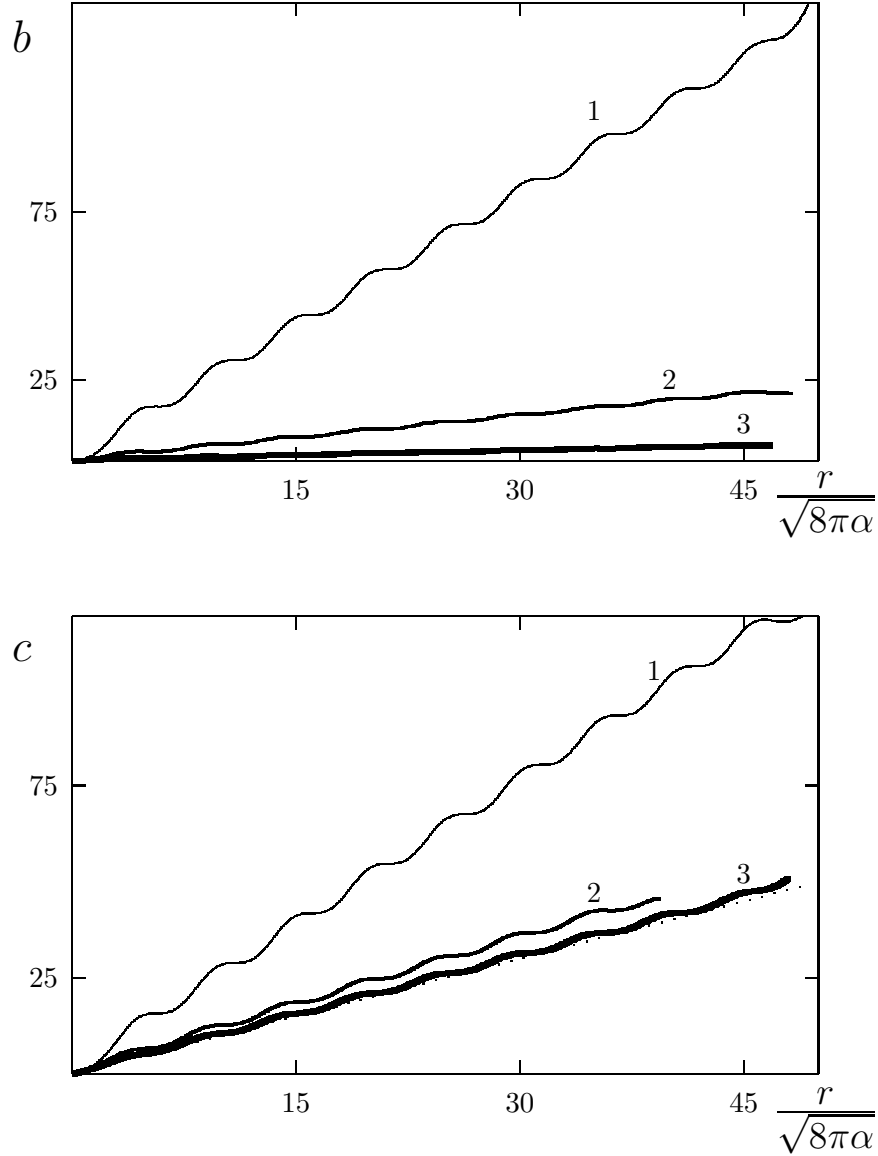
\setlength{\unitlength}{0.240900pt}


\vskip 5mm

Three cases which correspond to different boundary conditions are
represented on these figures:

1) $c_0 ^{\prime} =1; \quad c_0 ^{\prime \prime \prime} = 2; \quad \ \
b_0=1; \quad b_0 ^{\prime \prime} \approx 1.64218$

2) $c_0 ^{\prime} =1; \quad c_0 ^{\prime \prime \prime} = 1; \quad \ \
b_0=1; \quad b_0 ^{\prime \prime} \approx  0.69542$

3)  $c_0 ^{\prime} =1; \quad c_0 ^{\prime \prime \prime} = 0.5; \quad
b_0=1;\quad  b_0 ^{\prime \prime} \approx  0.29978$

The  asymptotic behaviour of $b(r)$ and $c(r)$ can be investigated
analytically. From the  numerical calculations, we see that for sufficiently
large $r$, both the functions can be represented as a sum
of two distinct components
\be
b(r)=b_{mon}(r) + \delta b(r); \quad c(r)=c_{mon}(r) + \delta c(r),
\ee
where $b_{mon}(r)$ and $c_{mon}(r)$ are strictly monotone increasing,
$\delta b(r)$ and $\delta c(r)$ are bounded oscillating functions.
The magnitudes of these components are connected as follows:
\be
\frac{\delta b(r)}{b_{mon}(r)}\sim\frac{\delta c(r)}
{c_{mon}(r)}\sim \epsilon<\!\!<1.
\ee
Moreover, the magnitudes of the
amplitudes and frequencies of $\delta b(r)$ and $\delta c(r)$ are
of order unity:
\be
\delta b^{(n)}(r)\sim\delta c^{(n)}(r)\sim \epsilon^0,
\ee
while
\be
b_{mon}(r)
\sim c_{mon}(r)\sim\epsilon^{-1}.
\ee
In fact, one supposes
that the function $b_{mon}(r)$ and $c_{mon}(r)$ increases like $r$
The numerical calculations also allow
us to estimate the magnitudes of the derivatives
of the monotonic components:
\be
b_{mon}^{(n)}(r)\sim c_{mon}^{(n)}(r)\sim \epsilon^{(n-1)}.
\ee

Having these basic estimates we can expand the equations
(\ref{eq1},\ref{eq2})
keeping the leading $O(\epsilon)$ terms
\be
\frac{\delta b^{\prime \prime \prime \prime}}{b_{mon}}+
\frac{1}{8 \pi \alpha} \left(\frac{\delta b^{ \prime \prime}}{2 b_{mon}}+
\frac{\delta c^{\prime \prime}}{c_{mon}} \right)=0,
\ee
\be
\frac{\delta c^{\prime \prime \prime \prime}}{c_{mon}}+
\frac{1}{8 \pi \alpha} \left(\frac{\delta b^{ \prime \prime}}{2 b_{mon}}+
\frac{\delta c^{\prime \prime}}{c_{mon}} \right)=0.
\ee
The solutions of these equations are
\be \label{db}
\frac{\delta b}{b_{mon}}=\frac{D_1}{\omega^2}e^{i \omega r}+
\frac{D_2}{\omega^2}e^{-i \omega r}+
\frac{D_3}{(\omega^2/3)} e^{(\omega/\sqrt{3})r}+
\frac{D_4}{(\omega^2 / 3)} e^{-(\omega/\sqrt{3})r}+
D_5 r + D_6,
\ee
\be \label{dc}
\frac{\delta c}{c_{mon}}=\frac{D_1}{\omega^2}e^{i \omega r}+
\frac{D_2}{\omega^2}e^{-i \omega r}-
\frac{D_3}{(\omega^2/3)} e^{(\omega/\sqrt{3})r}-
\frac{D_4}{(\omega^2/3)} e^{-(\omega/\sqrt{3})r}+
D_7 r + D_8,
\ee
where
\be
\displaystyle \omega^2=\frac{3}{16 \pi \alpha }
\ee
and
$D_i$ are arbitrary constants ($D_1$ and $D_2$ are complex conjugate,
$D_3, D_4, D_5, D_6, D_7, D_8$ are real).
Note that the parameters $D_i$
are constants only in the framework of the $O(\epsilon^2)$ approximation.
In fact, they may be slowly varying functions of $r$ such that their
derivatives are of $O(\epsilon^2)$. Note also that the boundary
conditions were chosen so that $D_3=0$.

In order to obtain the approximate but analytic expressions for the
monotonic components $b_{mon}(r)$ and $c_{mon}(r)$ more work is necessary.
For this we must expand equations (\ref{a1}-\ref{a4}) up
to order $O(\epsilon^2)$ and  substitute the solutions (\ref{db}, \ref{dc})
into these equations. Take into account
that these equations contain terms of three different types:
the terms which consist only of the monotonic functions $b_{mon}(r)$ and
$c_{mon}(r)$; the terms which contain one oscillating function
$\delta b(r)$ or $\delta c(l)$ (or its derivatives); and the terms
which contain a product of two oscillating functions $\delta b(r)$ or
$\delta c(l)$ (or its derivatives). After substituting the solutions
(\ref{db},\ref{dc}) into (\ref{a1}-\ref{a4}) one can check that
the these equations consist of two independent
parts: the monotonic one and the oscillating one.
Both these parts must be equated to zero separately.
Equating the monotonic part to zero we obtain the equations for
$b_{mon}(r)$ and $c_{mon}(r)$:

the $00$ component
\be \label{m1}
\frac{b_{mon}^{\prime \prime}}{b_{mon}}+
\frac{c_{mon}^{\prime \prime}}{c_{mon}}+
\frac{b_{mon}^{\prime} c_{mon} ^{\prime}} {b_{mon} c_{mon}}=
8 \pi \alpha \frac{8}{3} D_1 D_2 ,
\ee

the $zz$ component
\be \label{m2}
\frac{c_{mon}^{\prime \prime}}{c_{mon}}=0 ,
\ee

the $rr$ component
\be \label{m3}
\frac{b_{mon}^{\prime} c_{mon}^{\prime}}{b_{mon} c_{mon}}=
8 \pi \alpha \frac{8}{3} D_1 D_2 ,
\ee

the $\varphi \varphi$ component
\be \label{m4}
\frac{b_{mon}^{\prime \prime}}{b_{mon}}=0.
\ee
And the oscillating part of these equations gives the
correction of the order $\epsilon^2$ for $\delta b(r)$ and $\delta c(r)$.
The solutions of equations (\ref{m1}-\ref{m4}) are
\be
b=k_1 \frac{r}{\sqrt{8 \pi \alpha}} + k_2 + 0(\epsilon),
\ee
\be
c=k_3 \frac{r}{\sqrt{8 \pi \alpha}} + k_4 + 0(\epsilon),
\ee
\be
D_1 D_2 = \frac{3}{(8 \pi \alpha) 8 r^2} + 0(\epsilon ^3),
\ee
where $k_i$ are constants.

Finally, the asymptotic form of metric is
$$
ds^2=-dt^2+(k_1 \frac{r}{\sqrt{8 \pi \alpha}}+k_2)^2 \Big[1+
\frac{\sqrt{16 \pi \alpha / 3}}{r}
\sin \Big( \frac{r}{\sqrt{16 \pi \alpha / 3}}+r_0 \Big) \Big]^2 dz^2
$$
\be
+dr^2+(k_3 \frac{r}{\sqrt{8 \pi \alpha}}+k_4)^2 \Big[1+
\frac{\sqrt{16 \pi \alpha / 3}}{r}
\sin \Big( \frac{r}{\sqrt{16 \pi \alpha / 3}}
+r_0 \Big) \Big]^2 d \varphi ^2,
\ee
where $ \displaystyle \sin(r_0)= \frac{Re D_1}{\sqrt{D_1 D_2}}, \quad
\cos(r_0)=\frac{-Re D_1}{\sqrt{D_1 D_2}}$.
\vspace{0.5cm}

The asymptotic form of the angular part of these solutions
coincides with the angular part of cosmic string. However, in the
case of string the conic part of spacetime is
2-dimensional Minkowski spacetime minus a wedge and in the considered cases
the conic part of spacetime is 2-dimensional Minkowski
spacetime plus a wedge. The value of the cone excess angle is
of order unit and it is determined by the boundary conditions at the
symmetry axis.

\vskip 12pt

The work was supported by Russian Foundation
for Basic Researches grant No. 96-02-17366a

\section*{Appendix}

The equation $G_0^0 = 8\pi \langle T_0^0 \rangle$ give
$$
\frac{b^{\prime \prime }}b+\frac{c^{\prime \prime }}c+\frac{b^{\prime
}c^{\prime }}{bc}=8\pi \alpha \left( -\frac{2b^{\prime \prime \prime \prime }
}{3b}-\frac{2c^{\prime \prime \prime \prime }}{3c}+\frac{2b^{\prime \prime
\prime }b^{\prime }}{3b^2}-\frac{4b^{\prime \prime \prime }c^{\prime }}{3bc}-
\frac{4c^{\prime \prime \prime }b^{\prime }}{3bc}\right.
$$
$$
+\frac{2c^{\prime \prime
\prime }c^{\prime }}{3c^2}+\frac{4b^{\prime \prime 2}}{3b^2}-2\frac{%
b^{\prime \prime }c^{\prime \prime }}{bc}
+\frac{4c^{\prime \prime 2}}{3c^2}-\frac{2b^{\prime \prime }b^{\prime
2}}{3b^3}+\frac{4b^{\prime \prime }b^{\prime }c^{\prime }}{3b^2c}+\frac{%
2b^{\prime \prime }c^{\prime 2}}{3bc^2}+\frac{2c^{\prime \prime }b^{\prime 2}
}{3b^2c}
$$
\begin{equation}
\label{a1}
\left. +\frac{4c^{\prime \prime }b^{\prime }c^{\prime }}{3bc^2}-\frac{%
2c^{\prime \prime }c^{\prime 2}}{3c^3}-\frac{2b^{\prime 3}c^{\prime }}{3b^3c}%
+\frac{2b^{\prime 2}c^{\prime 2}}{3b^2c^2}-\frac{2b^{\prime }c^{\prime 3}}{%
3bc^3}\right).
\end{equation}

The equation $G_z^z = 8\pi \langle T_z^z \rangle$ give
$$
\frac{c^{\prime \prime }}c=8\pi \alpha \left( -\frac{4b^{\prime \prime
\prime \prime }}{3b}+\frac{2c^{\prime \prime \prime \prime }}{3c}+\frac{%
8b^{\prime \prime \prime }b^{\prime }}{3b^2}-\frac{8b^{\prime \prime \prime
}c^{\prime }}{3bc}+\frac{2c^{\prime \prime \prime }b^{\prime }}{3bc}-\frac{%
2c^{\prime \prime \prime }c^{\prime }}{3c^2}+2\frac{b^{\prime \prime 2}}{b^2}%
-\frac{2b^{\prime \prime }c^{\prime \prime }}{3bc}-\frac{4c^{\prime \prime }
}{3c^2}\right.
$$

\begin{equation}
\label{a2}
\left. -\frac{8b^{\prime \prime }b^{\prime 2}}{3b^3}+\frac{2b^{\prime \prime
}b^{\prime }c^{\prime }}{3b^2c}+\frac{4b^{\prime \prime }c^{\prime 2}}{3bc^2}%
-\frac{4c^{\prime \prime }b^{\prime 2}}{3b^2c}+\frac{8c^{\prime \prime
}b^{\prime }c^{\prime }}{3bc^2}+\frac{2c^{\prime \prime }c^{\prime 2}}{3c^3}+
\frac{4b^{\prime 3}c^{\prime }}{3b^3c}-\frac{2b^{\prime 2}c^{\prime 2}}{%
3b^2c^2}-\frac{4b^{\prime }c^{\prime 3}}{3bc^3}\right).
\end{equation}

The equation $G_r^r = 8\pi \langle T_r^r \rangle$ give
$$
\frac{b^{\prime }c^{\prime }}{bc}=8\pi \alpha \left( -\frac{4b^{\prime
\prime \prime }b^{\prime }}{3b^2}+\frac{2b^{\prime \prime \prime }c^{\prime }
}{3bc}+\frac{2c^{\prime \prime \prime }b^{\prime }}{3bc}-\frac{4c^{\prime
\prime \prime }c^{\prime }}{3c^2}+\frac{2b^{\prime \prime 2}}{3b^2}-\frac{%
2b^{\prime \prime }c^{\prime \prime }}{3bc}+\frac{2c^{\prime \prime 2}}{3c^2}%
+\frac{4b^{\prime \prime }b^{\prime 2}}{3b^3}\right.
$$

\be
\label{a3}
\left. -2\frac{b^{\prime \prime }b^{\prime }c^{\prime }}{b^2c}+
\frac{2b^{\prime
\prime }c^{\prime 2}}{3bc^2}+\frac{2c^{\prime \prime }b^{\prime 2}}{3b^2c}-2
\frac{c^{\prime \prime }b^{\prime }c^{\prime }}{3bc^2}+\frac{4c^{\prime
\prime }c^{\prime 2}}{3c^3}-\frac{2b^{\prime 3}c}{3b^3c}+\frac{2b^{\prime
2}c^{\prime 2}}{3b^2c^2}-\frac{2b^{\prime }c^{\prime 3}}{3bc^3} \right).
\ee

The equation $G_{\varphi}^{\varphi} = 8\pi \langle T_{\varphi}^{\varphi}
\rangle$ give
$$
\frac{b^{\prime \prime }}b=8\pi \alpha \left( \frac{2b^{\prime \prime \prime
\prime }}{3b}-\frac{4c^{\prime \prime \prime \prime }}{3c}-\frac{2b^{\prime
\prime \prime }b^{\prime }}{3b^2}+\frac{2b^{\prime \prime \prime }c^{\prime }
}{3bc}-\frac{8c^{\prime \prime \prime }b^{\prime }}{3bc}+\frac{8c^{\prime
\prime \prime }c^{\prime }}{3c^2}-\frac{4b^{\prime \prime 2}}{3b^2}-\frac{%
2b^{\prime \prime }c^{\prime \prime }}{3bc}+2\frac{c^{\prime \prime }}{c^2}%
\right.
$$

\begin{equation}
\label{a4}
\left. +\frac{2b^{\prime \prime }b^{\prime 2}}{3b^3}+\frac{8b^{\prime \prime
}b^{\prime }c^{\prime }}{3b^2c}-\frac{4b^{\prime \prime }c^{\prime 2}}{3bc^2}%
+\frac{4c^{\prime \prime }b^{\prime 2}}{3b^2c}+\frac{2c^{\prime \prime
}b^{\prime }c^{\prime }}{3bc^2}-\frac{8c^{\prime \prime }c^{\prime 2}}{3c^3}-
\frac{4b^{\prime 3}c^{\prime }}{3b^3c}-\frac{2b^{\prime 2}c^{\prime 2}}{%
3b^2c^2}+\frac{4b^{\prime }c^{\prime 3}}{3bc^3}\right).
\end{equation}

\end{document}